\documentclass[final,5p,times,twocolumn,authoryear]{elsarticle}

\usepackage{amssymb}
\usepackage{amsmath,amsfonts}
\usepackage{graphicx}
\usepackage{algorithm}
\usepackage[noend]{algpseudocode}
\usepackage{xcolor}
\usepackage{booktabs}
\usepackage{float}
\usepackage[noend]{algpseudocode}

\usepackage{hyperref}

\journal{Computers \& Security}

\begin{document}

\begin{frontmatter}


\title{\textcolor{blue}{Automated and Explainable Denial of Service Analysis for AI-Driven Intrusion Detection Systems}}

\author[first]{Paul Badu Yakubu}
\ead{pbaduyakubu@fordham.edu}

\author[first]{Lesther Santana}
\ead{lsantanacarmona@fordham.edu}


\author[first]{Mohamed Rahouti\corref{cor1}}
\ead{mrahouti@fordham.edu}

\author[second]{Yufeng Xin}
\ead{yxin@renci.org}

\author[Third]{Abdellah Chehri}
\ead{chehri@rmc.ca}

\author[Fourth]{Mohammed Aledhari}
\ead{Mohammed.Aledhari@unt.edu}

\cortext[cor1]{Corresponding author: Abdellah Chehri (chehri@rmc.ca)}

\affiliation[first]{organization={Computer and Information Science Department, Fordham University}, 
            city={New York},
            state={NY},
            country={USA}}

\affiliation[third]{organization={RENCI, University of North Carolina at Chapel Hill}, 
            city={Chapel Hill},
            state={NC},
            country={USA}}

\affiliation[Third]{organization={Department of Mathematics and Computer Science, Royal Military College of Canada}, 
            city={Kingston},
            state={ON},
            country={Canada}}

\affiliation[Fourth]{organization={Data Science Department, University of North Texas}, 
            city={Denton},
            state={TX},
            country={USA}}

\begin{abstract}
With the increasing frequency and sophistication of Distributed Denial of Service (DDoS) attacks, it has become critical to develop more efficient and interpretable detection methods. Traditional detection systems often struggle with scalability and transparency, hindering real-time response and understanding of attack vectors. This paper presents an automated framework for detecting and interpreting DDoS attacks using machine learning (ML). The proposed method leverages the Tree-based Pipeline Optimization Tool (TPOT) to automate the selection and optimization of ML models and features, reducing the need for manual experimentation. SHapley Additive exPlanations (SHAP) is incorporated to enhance model interpretability, providing detailed insights into the contribution of individual features to the detection process. By combining TPOT's automated pipeline selection with SHAP’s interpretability, this approach improves the accuracy and transparency of DDoS detection. Experimental results demonstrate that key features such as mean backward packet length and minimum forward packet header length are critical in detecting DDoS attacks, offering a scalable and explainable cybersecurity solution.
\end{abstract}


\end{frontmatter}

\section{Introduction} \label{sec:introduction}

Denial-of-Service (DoS) attacks are among the most persistent and significant threats to the availability and reliability of online services \cite{owusu2025unified}. By overwhelming systems with excessive traffic, these attacks can incapacitate critical infrastructure, leaving legitimate users unable to access vital resources. Given the widespread reliance on network services, the impact of such attacks can be severe, ranging from financial losses to compromised user trust. As the sophistication and frequency of DoS and Distributed Denial-of-Service (DDoS) attacks continue to evolve, the development of robust detection and mitigation mechanisms remains a top priority for cybersecurity professionals \cite{li2023comprehensive}.

The complexity of modern DDoS attacks lies in their ability to mimic legitimate traffic patterns, making it increasingly difficult for traditional rule-based systems to detect them. This challenge has led to the widespread adoption of machine learning(ML)-based approaches, which can analyze vast amounts of network traffic data and identify subtle anomalies indicative of an attack \cite{garcia2004anomaly, jones2005comparison}. However, while ML models offer improved detection capabilities, they come with their own set of challenges, such as the need for high-dimensional feature spaces, which can lead to increased computational overhead and potential overfitting \cite{al2023ddos}.

To address these challenges, feature selection and dimensionality reduction techniques are often employed to balance the trade-offs between accuracy and computational efficiency \cite{shukla2024iot}. These methods enable the identification of the most informative features—such as packet size, flow duration, and backward packet lengths—while minimizing noise and redundant data \cite{malik2023feature, guyon2003introduction}. Recent advancements in explainable AI have further enhanced the interpretability of these models, providing insights into how individual features contribute to the model's decisions and enabling more transparent cybersecurity frameworks \cite{lundberg2017unified, kalutharage2023explainable}.

\begin{figure*}[h]
  \centering
  \includegraphics[width=0.8\textwidth, keepaspectratio]{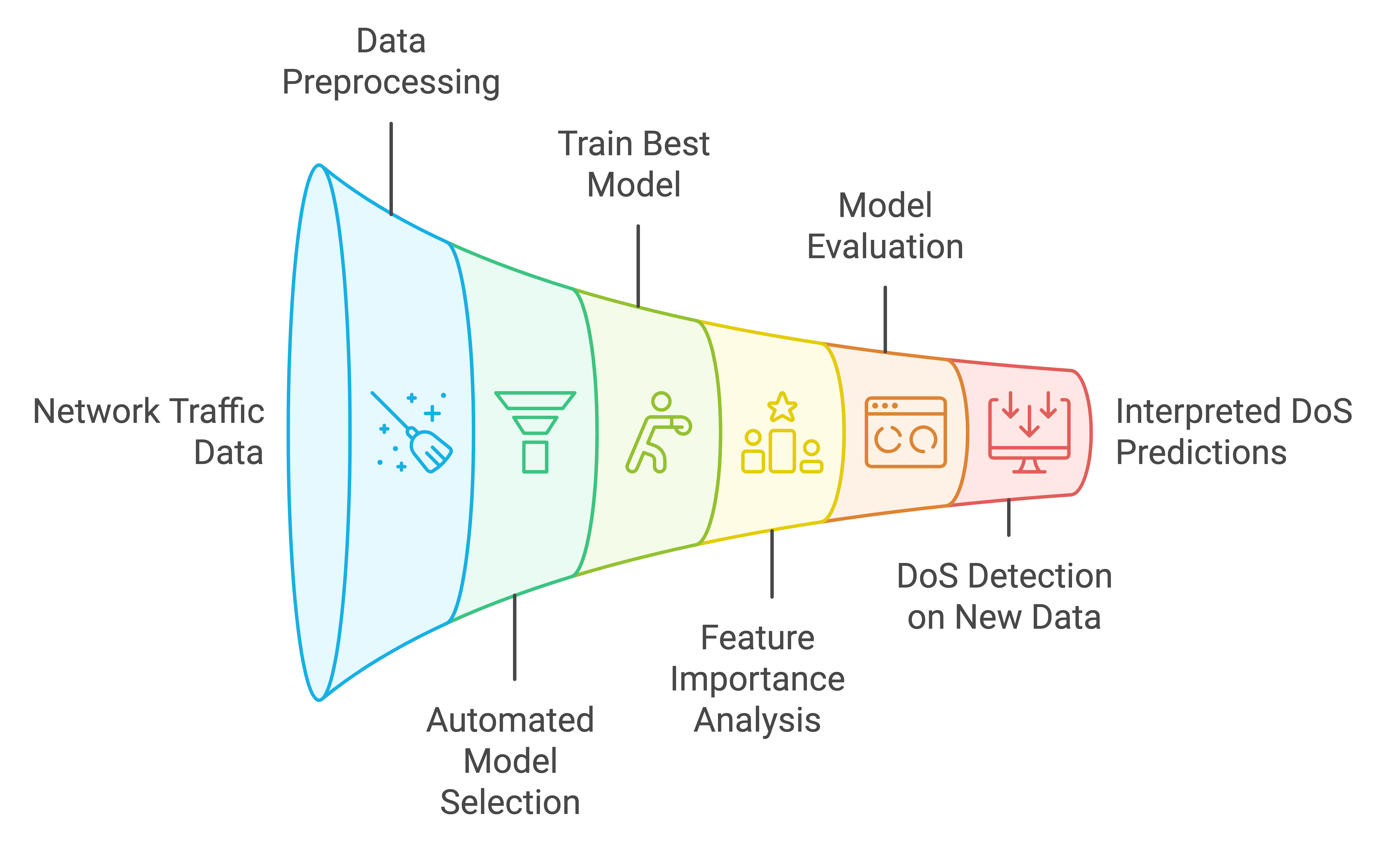}
  \caption{\textcolor{black}{Automated DoS detection pipeline.}}
  \label{fig:pipeline}
\end{figure*}

In this context, the proposed work leverages both automated ML tools and explainability techniques to build a robust DDoS detection framework. By combining the Tree-based Pipeline Optimization Tool (TPOT) for automating the selection and optimization of ML pipelines with SHapley Additive exPlanations (SHAP) for interpretability, this approach aims to improve both the accuracy and transparency of DDoS detection systems \cite{OlsonGECCO2016, lundberg2017unified}. Figure \ref{fig:pipeline} presents the automated DoS detection pipeline, which begins with raw network traffic data collection, followed by preprocessing and feature engineering steps. TPOT is utilized to automatically select the most suitable ML model, and SHAP is applied for explainability, providing insights into the model's decision-making process. This integration of automated pipeline optimization and interpretability not only enhances detection accuracy but also ensures that the system's decisions can be understood and trusted. As DDoS attacks become increasingly sophisticated, the use of such automated and interpretable solutions is essential to staying ahead of emerging threats.

\textcolor{blue}{The main contributions of this paper are summarized as follows:}
\begin{itemize}
    \item \textcolor{blue}{Leverages TPOT to automate the selection and optimization of ML models, significantly enhancing the robustness and accuracy of DoS attack detection by systematically identifying high-performing pipelines and optimal feature sets. This automation not only reduces the need for manual experimentation but also leads to more consistent and reproducible detection outcomes.}
    \item Integrates SHAP to provide detailed interpretability of the ML model, offering insights into the contribution of individual features to the model's predictions, enhancing transparency in cybersecurity applications.
    \item \textcolor{blue}{Proposes a framework demonstrating superior accuracy and effectiveness in detecting DoS attacks by employing advanced feature selection and dimensionality reduction techniques. Owing to its automated and model-agnostic design, the methodology is inherently flexible and can be extended to other cybersecurity threat detection tasks beyond DoS.}
\end{itemize}

The rest of this paper is organized as follows. Section \ref{sec:related} provides an overview of the background and related work, discussing traditional and ML-based approaches for DoS detection. Section \ref{sec:problem} presents the research problem and goals, detailing the specific challenges addressed in this study. In Section \ref{sec:methodology}, the methodology is described, outlining the integration of TPOT and SHAP for automated model selection and feature importance analysis. The results of the experimental evaluation are presented and analyzed in Section \ref{sec:evaluation}. \textcolor{blue}{Section \ref{sec:discussion} highlights the key lessons learned from this study and outlines future research directions.} Finally, Section \ref{sec:conclusion} concludes the paper and suggests directions for future research.

\section{Background and Related Work} \label{sec:related}

DoS and DDoS attacks represent some of the most significant threats to the availability and reliability of online services. These attacks flood targeted systems with an overwhelming amount of malicious traffic, rendering them inaccessible to legitimate users. Effective detection and mitigation of DoS and DDoS attacks are crucial for maintaining the integrity of networked systems. The dynamic and complex nature of these attacks, coupled with the increasing sophistication of attack methodologies, necessitates advanced detection mechanisms that can adapt in real-time \cite{fenil2020survey}.

\subsection{Traditional Methods for DoS and DDoS Detection}
Traditional methods for detecting DoS and DDoS attacks often rely on predefined signatures or anomaly-based detection techniques. Signature-based methods, while effective against known attacks, struggle to identify novel or evolving threats \cite{yang2022systematic}. Anomaly-based techniques, on the other hand, monitor network traffic for deviations from established patterns but can suffer from high false positive rates \cite{jones2005comparison}. 

\subsection{ML Approaches}

\textcolor{blue}{In recent years, ML has emerged as a powerful tool for detecting DoS and DDoS attacks, with numerous studies proposing tailored solutions leveraging supervised learning techniques. For example, Deepa et al.~\cite{deepa2018enhanced} introduced a hybrid detection system combining Support Vector Machines (SVM) with feature selection techniques to reduce false positives in DDoS detection. Their system achieved improved accuracy by selecting critical traffic features before classification, though it required extensive manual tuning.}

\textcolor{blue}{Similarly, Zhang and Zhou~\cite{zhang2008detection} utilized k-Nearest Neighbors (k-NN) with wavelet-based preprocessing to capture temporal patterns in packet-level data. While their method demonstrated high accuracy for specific attack types, it lacked generalization to newer, low-rate DDoS variants. Chen et al.~\cite{chen2006decision} explored decision tree learning for attack classification, emphasizing the interpretability and transparency of the resulting models. However, their approach did not incorporate feature selection or optimization, limiting its scalability to large datasets.}

\textcolor{blue}{Other studies such as Gupta et al.~\cite{gupta2018machine} developed a comprehensive ML-based DDoS mitigation framework for UAV networks using software-defined networking (SDN). Their system integrated multiple classifiers—Random Forests, Logistic Regression, Decision Trees, and XGBoost—into an ensemble, significantly improving traffic rerouting and attack mitigation performance. This framework highlighted the advantages of ensemble learning in capturing diverse traffic characteristics but required careful coordination between components. Another approach was proposed by Yang et al.~\cite{yang2014entropy}, who combined entropy analysis with ensemble classifiers to detect low-rate DDoS attacks. By monitoring statistical shifts in traffic entropy, their method could identify stealthy attacks that evade conventional detection; however, its reliance on entropy thresholds reduced adaptability to dynamically evolving threats.}

\textcolor{blue}{Further advances have introduced increasingly sophisticated ML strategies tailored for DoS and DDoS detection in diverse network environments. Zhao et al.~\cite{zhao2023enhancing} emphasized the importance of feature engineering to improve detection performance, especially in the context of evolving attack behaviors. Santana et al.~\cite{santana2024exploring} further explored explainability techniques to highlight feature importance and enhance trust in ML-based DoS detectors. For resource-constrained settings such as the Internet of Medical Things (IoMT), Clark et al.~\cite{clark2024machine} presented a lightweight detection framework optimized for edge devices, while Hafid et al.~\cite{hafid2025optimizing} proposed a cost-aware and interpretable ML system tailored for IoMT-based intrusion detection. To improve data efficiency, Alfatemi et al.~\cite{alfatemi2025protomaml} introduced ProtoMAML, a hybrid meta-learning framework that combines Prototypical Networks with MAML for effective DDoS detection under limited training data. Owusu et al.~\cite{owusu2025robust} proposed a robust detection framework using combinatorial fusion and generative AI, and their subsequent work~\cite{owusu2024online} explored online DDoS detection using adaptive sampling and change point detection. Finally, Al-Zewairi et al.~\cite{al2025multi} developed a multi-stage intrusion detection system leveraging enhanced Zero Trust principles to detect unknown and emerging threats in both IoT and traditional network environments. These contributions collectively reflect the ongoing shift toward adaptive, interpretable, and resource-efficient ML-based intrusion detection systems.}

\textcolor{blue}{In summary, while traditional ML approaches have demonstrated efficacy in DDoS detection, they often involve manual pipeline construction and lack interpretability. These limitations motivate the need for automated and explainable detection frameworks, such as the one proposed in this work, which leverages TPOT for pipeline optimization and SHAP for feature-level interpretability.}

\subsection{Feature Selection and Dimensionality Reduction}

\textcolor{black}{Feature selection is a critical step in enhancing the performance of ML models for DoS detection. High-dimensional feature spaces provide detailed insights into network behavior but also introduce challenges such as increased computational complexity and the risk of overfitting. Techniques such as Principal Component Analysis (PCA) and other dimensionality reduction methods transform features into a lower-dimensional space while preserving essential information \cite{guyon2003introduction}. Recent studies have highlighted the effectiveness of feature engineering and selection in detecting DDoS attacks in the context of IoT, where high-dimensional data often leads to inefficiencies. Malik et al. \cite{malik2023feature} demonstrated that a comprehensive feature engineering approach significantly improves detection accuracy by reducing irrelevant features while maintaining critical information. Furthermore, Abu Bakar et al. \cite{abu2023intelligent} proposed an intelligent agent-based detection system that combines automatic feature extraction and selection, leading to enhanced performance in identifying DDoS attacks with minimal manual intervention.}

\textcolor{black}{The importance of distinguishing between feature selection and feature extraction has also been discussed by Ngo et al. \cite{ngo2024machine}, who compared the two methods in the context of intrusion detection. Their findings showed that while feature selection helps in retaining the most important features, feature extraction, such as PCA, is effective in reducing feature dimensions, resulting in improved model performance without compromising accuracy. Additionally, Kemp et al. \cite{kemp2023approach} applied feature reduction to effectively detect application-layer DoS attacks, emphasizing that dimensionality reduction techniques can minimize computational overhead, making real-time detection feasible even in resource-constrained environments.}

\subsection{Interpretability and Explainability in ML Models}

\textcolor{black}{As ML models become more complex, understanding their decision-making processes becomes increasingly challenging. This has led to a growing focus on interpretability and explainability in ML. Methods such as SHAP provide interpretable insights into feature contributions and model predictions. SHAP values help quantify the impact of each feature on the model's output, offering a clearer understanding of how predictions are made \cite{lundberg2017unified}. Kalutharage et al. \cite{kalutharage2023explainable} explored an explainable AI (XAI) approach to identify DDoS attacks in IoT networks, using explainable methods to provide insights into the decision-making of deep learning models, which is critical for gaining trust in automated systems.}

\textcolor{black}{The importance of explainability is further underscored by Sharma et al. \cite{sharma2024explainable}, who introduced an XAI-based deep learning approach for intrusion detection in IoT networks, highlighting that explainability not only improves trust but also aids in identifying potential model biases. Alzu'bi et al. \cite{alzu2024explainable} similarly leveraged deep transfer learning with XAI techniques for classifying DDoS attacks, allowing for detailed analysis of feature importance and enhancing the interpretability of otherwise black-box models. Moreover, Arreche et al. \cite{arreche2024xai} evaluated several black-box XAI frameworks for network intrusion detection, emphasizing the need for effective XAI tools that can elucidate the underlying reasoning of complex ML models.}

\textcolor{black}{The review by Nwakanma et al. \cite{nwakanma2023explainable} on the application of XAI in connected vehicles and other intelligent systems underscores the growing trend towards integrating interpretability into cybersecurity solutions. By making models more transparent, stakeholders are better equipped to understand the risks associated with false positives or negatives. The analysis provided by Al-Shareeda et al. \cite{al2023ddos} further confirms the necessity of explainable models in the domain of DDoS detection, where complex attack patterns necessitate clear explanations of model decisions to ensure reliability and effectiveness in practical applications.}

\subsection{Uniqueness of This Paper}

\textcolor{black}{This paper uniquely integrates the TPOT and SHAP to automate model selection and enhance interpretability for DoS/DDoS attack detection. Unlike existing studies that often rely on manual feature selection or traditional ML approaches, this work automates the pipeline selection, thus reducing experimental overhead while optimizing detection accuracy and efficiency. Additionally, the incorporation of SHAP provides detailed insights into feature contributions, addressing the ``black box" nature of many ML models in cybersecurity. This combination of automation and interpretability significantly enhances both the effectiveness and trustworthiness of DDoS detection systems.}

\begin{algorithm}[h]
\caption{Automated DoS detection and feature engineering using TPOT and SHAP.}
\label{alg:alg1}
\begin{algorithmic}[1]
\State \textbf{Input}: Network traffic dataset $D = \{(X_{train}, y_{train}), (X_{test}, y_{test})\}$
\State \textbf{Output}: Trained model $M$ with selected features and interpretability analysis

\Procedure{Feature Selection and Model Training with TPOT}{}
    \State \textit{// Step 1: Configure TPOT for automated pipeline optimization}
    \State Initialize TPOT with generations $g$, population size $p$, and scoring metric $s$
    \State Set TPOT configuration to include feature selection methods
    \vspace{5pt}  
    \State \textit{// Step 2: Train and optimize the pipeline}
    \State Fit TPOT on training data $(X_{train}, y_{train})$
    \State Extract the best pipeline $P_{best}$ from TPOT
    \vspace{5pt}  
    \State \textit{// Step 3: Evaluate the optimized model}
    \State Evaluate $P_{best}$ on test data $(X_{test}, y_{test})$
    \State Save the trained model $M$ for further analysis
\EndProcedure

\Procedure{Feature Importance Analysis with SHAP}{}
    \State \textit{// Step 4: Compute SHAP values for the best model}
    \State Calculate SHAP values for features in $P_{best}$ using the test data
    \State Rank features based on their SHAP importance values
    \vspace{5pt}  
    \State \textit{// Step 5: Visualize and interpret feature importance}
    \State Generate SHAP summary plots for visual interpretation
    \State Analyze the impact of top-ranked features on the model's predictions
\EndProcedure

\Procedure{DoS Detection}{}
    \State \textbf{Input}: New network traffic data $X_{new}$
    \State \textit{// Step 6: Predict using the trained model}
    \State Apply the trained model $M$ to predict labels for $X_{new}$
    \State Output the predicted labels $\hat{y}_{new}$
\EndProcedure
\State \textbf{return} Trained model $M$, SHAP analysis results, and predictions $\hat{y}_{new}$
\end{algorithmic}
\end{algorithm}

\section{Research Problem and Goals} \label{sec:problem}

\textcolor{black}{The detection and mitigation of DoS and DDoS attacks remain a significant challenge in ensuring the availability and reliability of networked systems \cite{mirkovic2004ddos, garcia2004anomaly, li2020recent}. These attacks generate a high volume of malicious traffic, which overwhelms targeted systems, rendering them inaccessible to legitimate users. The complexity of detecting DDoS attacks lies in the high-dimensional nature of network traffic data, the need for real-time analysis, and the evolving tactics of attackers, which require adaptive and robust detection mechanisms \cite{behal2018ddos, li2018ddos}. Traditional approaches, such as signature-based and anomaly-based detection, face limitations, including the inability to effectively recognize novel and sophisticated attack patterns, high false positive rates, and the challenges associated with manual feature selection \cite{jones2005comparison, yang2014entropy}.}

\textcolor{black}{To address these challenges, this study aims to enhance DDoS detection by automating the selection of ML models and optimizing feature selection using TPOT. By automating these processes, the study seeks to reduce the manual experimentation required to identify effective models and feature sets, thereby improving scalability and efficiency \cite{OlsonGECCO2016}. Furthermore, the lack of interpretability in ML models for DDoS detection has been a critical barrier to their deployment, as stakeholders need to understand and trust the model's decisions. To address this issue, SHAP is utilized to provide transparency into the model's decision-making process, highlighting the contribution of individual features and improving trustworthiness \cite{lundberg2017unified}.}

\textcolor{black}{This research focuses on creating an automated and interpretable framework for DDoS detection that not only optimizes model selection and feature importance but also enhances transparency, making it a practical solution for real-world network security applications. The ultimate goal is to streamline the development of ML models for DDoS detection while ensuring their effectiveness, scalability, and transparency, thereby contributing valuable insights into the deployment of AI-driven cybersecurity solutions.}

\section{Methodology} \label{sec:methodology} 

\textcolor{black}{Algorithm \ref{alg:alg1} outlines our comprehensive approach to DoS detection by integrating automated ML and interpretability techniques. It begins with configuring and utilizing the TPOT to automatically select the optimal feature subset and ML pipeline, ensuring enhanced accuracy and efficiency. Once the best pipeline is identified, the model's feature importance is further analyzed using SHAP, which ranks features based on their contribution to the model's predictions, providing critical insights into the model's decision-making process. The final model, equipped with optimized features and interpretability, is then employed to predict DoS attacks on new network traffic data, offering a robust and transparent solution for AI-driven cybersecurity.}

In this study, as discussed above, we leverage the TPOT alongside SHAP to enhance feature selection, model performance, and interpretability in ML models for detecting DoS attacks. TPOT automates the design of the ML pipeline, optimizing feature selection and hyperparameters, thereby identifying the best-performing model configurations. By streamlining the process of discovering optimal model architectures and settings, TPOT improves the accuracy and efficiency of DoS detection models. Complementing TPOT, SHAP provides detailed, interpretable insights into the contributions of individual features and the rationale behind model predictions. The integration of TPOT's automated optimization and SHAP's interpretability facilitates a comprehensive understanding of model behavior, making it possible to not only achieve high performance in detecting DoS attacks but also to understand the underlying factors influencing model decisions. This combination enhances both the performance and transparency of ML models in intrusion detection applications \cite{OlsonGECCO2016}.

\begin{table}[h]
\centering
\begin{tabular}{|p{2.5cm}|p{5.2cm}|} \hline
Feature & Info  \\ 
 \hline
Decision tuple & ID, src/dest IP, src/dst port, protocol \\ \hline
Time &  Timestamp, duration \\ \hline
Fwd pkts & Total, len (total, max, min, std, min) \\ \hline
Bwd pkts & Total, len (total, max, min, std, min)  \\ \hline
IAT & Mean, std, max, min \\ \hline
Fwd IAT & Total, mean, std, max, min \\ \hline
Bwd IAT & Total, mean, std, max, min \\ \hline
Fwd flags & Push, URG \\ \hline
Flags & Bwd (Push, URG), Count \\ \hline
Pkts len/size & Pkts (min, std, max, mean, var), size (avg) \\ \hline
Pkt loss & Down/up ratio \\ \hline
Flags count & FIN/SYN/RST/PSH/ACK/URG/ CWE/ECE   \\ \hline
Fwd pkt header & Len, avg (Seg size, bytes/bulk, bulk rate) \\ \hline
Bwd pkt header & Len, avg (Seg size, bytes/bulk, bulk rate)     \\ \hline
Subflow & Fwd/bwd (pkts, Bytes) \\ \hline
Init win Bytes & Fwd, Bwd    \\ \hline
Active/idle & Mean, max, std, min  \\ \hline
Other labels & Inbound, Similar HTTP \\ \hline
\end{tabular}
\caption{\textcolor{black}{Flow features and statistics in the Lycos-IDS2017 dataset \cite{sharafaldin2018ids}.}}
\label{tab:data-features}
\end{table}

\subsection{\textcolor{black}{Dataset}}

\textcolor{black}{The Lycos-IDS2017 dataset \cite{sharafaldin2018ids} we used in this paper is a comprehensive network traffic dataset that includes various types of network attacks, including DDoS. The dataset was preprocessed to handle missing values, encode categorical variables, and normalize continuous features. We focus on a subset of features most relevant to DDoS detection, as identified by the TPOT optimization process \cite{sharafaldin2018ids}.}

\textcolor{black}{Key attributes of the dataset listed in Table \ref{tab:data-features} include \texttt{flag\_rst} (Reset flag in TCP headers), \texttt{pk\_len\_std} (standard deviation of packet lengths), \texttt{fwd\_subflow\_bytes\_mean} (average byte size of forward subflows), \texttt{flow\_duration} (total duration of network flows), \texttt{bwd\_pkt\_len\_mean} (average size of backward packets), and\\ \texttt{bwd\_pkt\_len\_tot} (total length of backward packets). These features provide an in-depth view of network behaviors, supporting our analysis effectively.}

\subsection{Automated Model Selection through TPOT}

TPOT is a framework that automates the design and optimization of ML pipelines through genetic programming. It systematically explores various combinations of preprocessing techniques, feature selection methods, and ML algorithms to identify the most effective pipeline for a given dataset. In this work, we outline the methodology for employing TPOT in the automated selection of models and features specifically for DoS detection (Algorithm \ref{alg:TPOT}). TPOT can be configured to include feature selection operators within the optimization process by incorporating desired feature selection methods. The TPOT pipeline is then fitted to the training data, during which TPOT iteratively evaluates and optimizes different pipelines that encompass feature selection steps. Upon completion of the optimization process, the best-performing pipeline is extracted, encapsulating the optimal feature selection techniques and associated parameters \cite{OlsonGECCO2016}.

TPOT optimizes a pipeline \( P \) by solving:

\[
P^* = \arg\max_{P \in \mathcal{P}} \text{Accuracy}(P(D_{\text{train}})),
\]

where \( P^* \) is the optimal pipeline, \( P \) is the set of all possible pipelines, and \( D_{\text{train}} \) is the training dataset. The pipeline consists of a sequence of data preprocessing steps, feature selection, and model selection.

\begin{algorithm}[h]
\caption{\textcolor{black}{Automated model selection using TPOT.}}
\label{alg:TPOT}
\begin{algorithmic}[1]
    \Require Training dataset $(X_{\text{train}}, y_{\text{train}})$, generations $g=2$, population size $p=30$, scoring metric $s=$ `accuracy', configuration $c=$ `TPOT light'
    \Ensure Optimized ML pipeline $\mathcal{P}_{\text{best}}$

    \State \textbf{Initialize} TPOT with parameters $(g, p, s, c)$
    \State \textbf{Start timer} to measure total runtime
    \State \textbf{Fit} TPOT on $(X_{\text{train}}, y_{\text{train}})$ to evolve pipelines
    \State \textbf{Obtain} best pipeline $\mathcal{P}_{\text{best}}$
    \State \textbf{Evaluate} accuracy of $\mathcal{P}_{\text{best}}$ on test data $(X_{\text{test}}, y_{\text{test}})$
    \State \textbf{Print} accuracy and total runtime
    \State \textbf{Export} $\mathcal{P}_{\text{best}}$ to a Python file
\end{algorithmic}
\end{algorithm}

In our work, we use TPOT to automate selecting the best ML pipeline using genetic programming. The key parameters we set for the TPOTClassifier are `generations=2', `population\_size=30', and `scoring=accuracy', which means the algorithm will evolve the model over 2 generations with a population size of 30 pipelines per generation, optimizing for accuracy. The `config\_dict=TPOT light' indicates a simplified configuration to speed up the process. The timer measures the total runtime for the optimization process, and the `tpot.fit(X\_train, y\_train)' command trains the model on the training data. After fitting, the model's accuracy on the testing data and the total time taken for the process are printed. Finally, the best model pipeline discovered by TPOT is exported to a Python file.

The output shows that after two generations, the TPOT algorithm identified the best pipeline as a DecisionTreeClassifier with specific parameters: `criterion=entropy', `max\_depth=10',\\ `min\_samples\_leaf=2', and `min\_samples\_split=7'. The best internal cross-validation score achieved was approximately 0.9992, and the best accuracy score on the test data was 0.9991, indicating an excellent model performance. The process took 11 minutes and 22 seconds, demonstrating TPOT's efficiency in automating the model selection and optimization process while achieving high accuracy. This is particularly useful for leveraging ML without extensive manual experimentation and tuning of models.

\subsection{Intelligent TPOT Pipeline-based Feature Importance Extraction}

\begin{table}[h]
\centering
\begin{tabular}{r|l|r}
\toprule
Rank & Feature & Importance \\
\midrule
1  & bwd\_pkt\_len\_mean       & 0.620557    \\
2  & fwd\_pkt\_hdr\_len\_min    & 0.117785    \\
3  & flag\_fin               & 0.0803743   \\
4  & fwd\_iat\_min            & 0.0378663   \\
5  & pkt\_len\_var            & 0.0349627   \\
6  & fwd\_iat\_mean           & 0.0317716   \\
7  & fwd\_tcp\_init\_win\_bytes & 0.0131679   \\
8  & fwd\_flag\_psh           & 0.00993987  \\
9  & down\_up\_ratio          & 0.00961987  \\
10 & fwd\_pkt\_cnt            & 0.00924045  \\
11 & bwd\_tcp\_init\_win\_bytes & 0.00877884  \\
12 & active\_min             & 0.00588071  \\
13 & bwd\_iat\_min            & 0.00492238  \\
14 & bwd\_iat\_mean           & 0.00411989  \\
15 & fwd\_pkt\_hdr\_len\_tot    & 0.00354573  \\
16 & bwd\_pkt\_per\_s          & 0.00197074  \\
17 & iat\_std                & 0.00171884  \\
18 & flag\_rst               & 0.00107015  \\
19 & fwd\_pkt\_len\_mean       & 0.000886202 \\
20 & bwd\_iat\_std            & 0.000457246 \\
21 & fwd\_iat\_max            & 0.00043774  \\
22 & fwd\_pkt\_len\_max        & 0.000408617 \\
23 & flag\_SYN               & 0.000408114 \\
24 & pkt\_per\_s              & 0.000110606 \\
\bottomrule
\end{tabular}
\caption{Feature importance in descending order.}
\label{tab:feature-importances}
\end{table}

Next, we extract the important features from the best pipeline in the TPOT file. Our focus was on the features with an importance score greater than zero because these features contribute meaningfully to the model's predictions, whereas features with an importance score of zero have no influence and can be disregarded. By filtering out non-influential features, we reduce the complexity of the model without compromising its performance, which is crucial for enhancing interpretability and avoiding overfitting. The results of this process are shown in Table \ref{tab:feature-importances}, which depicts the feature importances in descending order. The most important feature is \texttt{bwd\_pkt\_len\_mean} with an importance score of 0.620557, followed by \texttt{fwd\_pkt\_hdr\_len\_min} at 0.117785, and \texttt{flag\_fin} at 0.0803743. The table lists 24 features with non-zero importance scores, indicating their relevance to the Decision Tree model's performance. This provides insights into which features contribute most significantly to the model's decision-making process, helping to better understand and interpret the model's predictions. The detailed ranking of features allows for further analysis and potential feature selection for future modeling efforts.

In the context of detecting DDoS attacks, \texttt{bwd\_pkt\_len\_mean} (Mean Backward Packet Length) is identified as the most significant feature, with an importance score of 0.620557. This feature represents the average size of packets transmitted from the target server back to the client, which may be an attacker in the case of a DDoS scenario. During a DDoS attack, attackers typically inundate the target with numerous small and repetitive requests. Consequently, the server responds with a large number of uniform, small packets, resulting in a consistent mean backward packet length. The predictability and uniformity of the backward traffic generated in response to such attacks render \texttt{bwd\_pkt\_len\_mean} a critical indicator for identifying malicious activities. The prominence of this feature, compared to others, underscores the strong correlation between consistent response packet sizes and the occurrence of a DDoS attack.

The second most significant feature in detecting DDoS attacks is \texttt{fwd\_pkt\_hdr\_len\_min} (Minimum Forward Packet Header Length), with an importance score of 0.117785. This feature measures the smallest header size in the forward packets sent from the attacker to the server. DDoS attacks, particularly volumetric ones, often involve minimal headers to maximize the number of packets sent within a short timeframe. Attackers utilize minimal packet headers to evade detection while consuming minimal bandwidth, thereby overwhelming the target. The consistency in the small header sizes of these forwarded packets makes \texttt{fwd\_pkt\_hdr\_len\_min} an effective indicator of DDoS traffic, as normal network traffic typically exhibits more variability in packet header lengths.

In addition to the aforementioned features, \texttt{flag\_fin} and\\ \texttt{fwd\_iat\_min} are ranked third and fourth, respectively, in terms of their importance. \texttt{flag\_fin} (importance score: 0.0803743) represents the presence of the TCP FIN flag, which indicates the termination of a connection. Abnormal patterns in the use of this flag may suggest malicious attempts to establish and abruptly terminate numerous connections, which is a common behavior in DDoS attacks such as SYN floods or TCP-based attacks. Similarly, \texttt{fwd\_iat\_min} (importance score: 0.0378663), which measures the minimum inter-arrival time of forward packets, is crucial for distinguishing attack traffic from legitimate user behavior. DDoS attacks typically involve rapid bursts of packets with very short intervals between them, whereas normal traffic is characterized by more sporadic packet transmissions. This distinction makes \texttt{fwd\_iat\_min} a valuable feature in identifying high-frequency attacks.

Other features, such as `pkt\_len\_var` (Packet Length Variance) and `fwd\_iat\_mean` (Mean Forward Inter-Arrival Time), also play a significant role. The `pkt\_len\_var` (0.0349627) captures the variability in packet lengths, which tends to be lower in DDoS traffic due to the repetitive nature of attack packets. In contrast, normal traffic is more likely to show a higher variance in packet lengths, making this feature useful for distinguishing between attack and legitimate traffic. Similarly, `fwd\_iat\_mean` (0.0317716) measures the average time between forward packets and can help identify anomalies in packet timing, such as the consistent rapid-fire requests seen during DDoS attacks. These features, although lower in rank, still contribute significantly to identifying patterns specific to DDoS attacks by analyzing the consistency and timing characteristics of network traffic.

\subsection{Utilizing SHAP for Feature Importance Analysis Across Different Classes}

SHAP values are a method used to explain the output of any ML model by employing a game-theoretic approach that measures each feature's contribution to the final outcome. In ML, each feature is assigned an importance value representing its contribution to the model's output. SHAP values illustrate how each feature affects each final prediction, the relative significance of each feature, and the model's reliance on feature interactions. Features with positive SHAP values have a positive impact on the prediction, while those with negative values exert a negative influence. The magnitude of these values indicates the strength of each feature's effect.

Interpreting the model's decision-making process is crucial for understanding how and why certain predictions are made. In this study, SHAP was used to provide insights into the contributions of individual features to the model's predictions. SHAP values are derived from cooperative game theory, specifically the concept of Shapley values, which allocate credit for a cooperative game's outcome to individual players (in this case, features) \cite{NIPS2017_7062}.

For a given feature \( i \) in an instance \( x \), the SHAP value \( \phi_i \) is computed as:

\[
\phi_i = \sum_{S \subseteq F \setminus \{i\}} \frac{|S|! \, (|F| - |S| - 1)!}{|F|!} \left[ f(S \cup \{i\}) - f(S) \right]
\]

Where:
\begin{itemize}
    \item \( F \) is the set of all features,
    \item \( S \) is a subset of features excluding \( i \),
    \item \( f(S) \) is the model's prediction based only on features in \( S \).
\end{itemize}

This equation essentially calculates the marginal contribution of each feature to the prediction, averaged over all possible feature combinations.

In our final stage, we extract and display the feature importances from the best pipeline identified by TPOT and compare them with the important features obtained in the previous step. First, the final pipeline model is retrieved using \texttt{tpot.fitted\_pipeline\_}, and the column names of the input data are assigned to \texttt{feature\_names}. If the final model in the pipeline is a \texttt{DecisionTreeClassifier}, the feature importances are extracted, and a dictionary is created pairing each feature name with its corresponding importance. Features with zero importance are filtered out, and the remaining features are sorted by importance in descending order. The sorted features are then formatted into a pandas DataFrame, including a ranked index, and printed in a markdown table format.

\begin{figure}[h]
  \centering
  \includegraphics[width=0.6\textwidth, keepaspectratio]{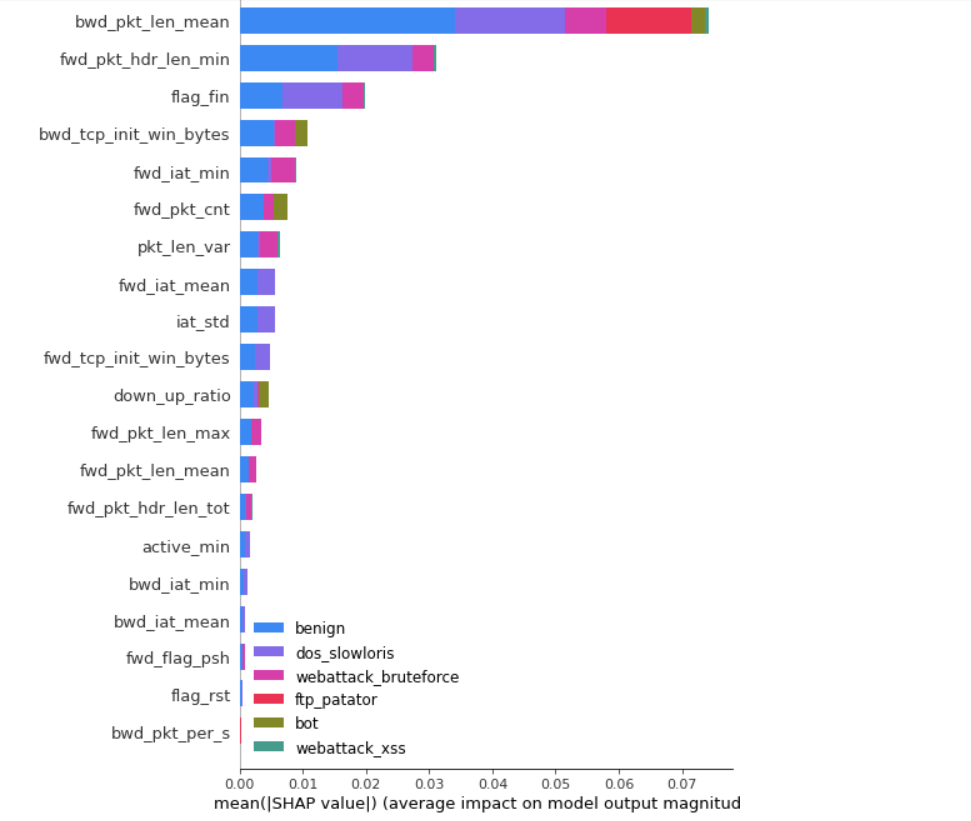}
  \caption{A summary plot showing important features for all classes}
  \label{fig:sample-image}
\end{figure}

Figure \ref{fig:sample-image} shows the SHAP additive values of the important features across all six labels, where the contribution of each feature to the final prediction can be computed independently and then summed up.

The top-ranked feature, bwd\_pkt\_len\_mean (Mean Backward Packet Length), is critical because it measures the average length of packets sent from the target server back to the source. During a DDoS attack, the server typically responds to an overwhelming number of small, repetitive requests from multiple sources. These requests are often minimal in nature, such as HTTP GET requests, which demand a simple acknowledgment or small response. Consequently, the size of the packets sent back (backward packets) tends to be consistent and relatively small. The mean backward packet length thus reveals an important signature: the more consistent and shorter the backward packets are, the more likely it is that the server is under attack. In normal traffic conditions, the size of backward packets would vary considerably based on the nature of the requests, such as loading complex pages or responding to various types of queries. A consistent pattern of small backward packet lengths can indicate the presence of a DDoS attack, where the server is constantly bombarded with repetitive requests, leading to a uniform response size \cite{behal2018ddos, das2019packet, amini2015feature}.

Similarly, `fwd\_pkt\_hdr\_len\_min` (Minimum Forward Packet Header Length) provides vital insights into the nature of the traffic sent to the server, specifically by measuring the smallest header size in the packets sent from the attackers to the target. In many DDoS attacks, especially volumetric ones, the attacker’s goal is to send as many packets as possible in the shortest time frame to overwhelm the server's resources. These attack packets often have minimal headers, containing just enough information to be processed but not enough to carry substantial or meaningful data. This results in a consistently small forward packet header size during the attack. In contrast, legitimate traffic typically has varying header sizes depending on the type of communication, as different requests and services require different amounts of header information. Therefore, a persistently low minimum packet header length can signal that a DDoS attack is underway, characterized by an unusually high volume of small, repetitive packets designed to strain the target’s resources. Together, `bwd\_pkt\_len\_mean` and `fwd\_pkt\_hdr\_len\_min` serve as critical indicators, helping to distinguish between normal network activity and the abnormal traffic patterns seen during DDoS attacks \cite{kumar2019ddos,li2018ddos}.

\begin{table}[h]
    \centering
    \begin{tabular}{|c|c|c|c|}
        \hline
        \textbf{Traffic Label} & \textbf{Label} & \textbf{Train set} & \textbf{Test set} \\
        \hline
        benign & 0 & 83163 & 194140 \\
        bot & 1 & 18 & 40 \\
        dos\_slowloris & 2 & 1045 & 2406 \\
        ftp\_patator & 3 & 579 & 1344 \\
        webattack\_bruteforce & 4 & 434 & 926 \\
        webattack\_xss & 5 & 36 & 121 \\
        \hline
    \end{tabular}
    \caption{Nteworking flow label distribution in train and test sets.}
    \label{tab:traffic_label_distribution}
\end{table}

\section{\textcolor{blue}{Performance Evaluation and Feature Impact Analysis}}
\label{sec:evaluation}

\begin{table}[h]
\footnotesize
    \centering
    \begin{tabular}{|c|c|c|c|c|c|c|}
    \hline
     & benign & bot & dos\_s.. & ftp\_p.. & web..\_b.. & web..\_x.. \\ \hline
    benign & 194089 & 1 & 7 & 9 & 9 & 4 \\ \hline
    bot & 4      & 39 & 0 & 0 & 0 & 0 \\ \hline
    dos\_s.. & 36     & 0  & 2397 & 0 & 0 & 0 \\ \hline
    ftp\_p.. & 0      & 0  & 0  & 1335 & 0 & 0 \\ \hline
    web..\_b .. & 11     & 0  & 2  & 0 & 917 & 117 \\ \hline
    web..\_x.. & 0      & 0  & 0  & 0 & 0 & 0 \\ \hline
    \end{tabular}
    \caption{Confusion matrix.}
    \label{tab:confusion-matrix}
\end{table}

In the evaluation of our model's performance, the confusion matrix in Table \ref{tab:confusion-matrix} provides a detailed breakdown of correct and incorrect predictions. The matrix shows that the model accurately detects most benign traffic, with very few false positives. However, the small number of misclassifications in other attack types highlights potential areas for improvement, particularly in identifying web-based attacks such as XSS (Cross-Site Scripting) and FTP-based attacks. We calculated the false positive rate (FPR) and false negative rate (FNR) to gain further insights. For example, the FPR for benign traffic was 0.002\%, indicating the model's robustness in detecting legitimate traffic. Meanwhile, the FNR for webattack\_xss was relatively high at 3.7\%, suggesting a need for model refinement in handling specific attack vectors.

In addition to the AUC-ROC, we used a Precision-Recall Curve (PRC) to assess the model's performance, particularly due to the imbalanced nature of the dataset where true negatives (benign traffic) dominate. The PRC provides a better understanding of the model's ability to correctly identify attacks. As depicted in Figure \ref{fig:sample-image}, the model achieves a high precision across most attack categories, reflecting its strong capability to minimize false positives. However, the recall for the webattack\_xss category was lower, indicating that while the model rarely misclassifies benign traffic as malicious, it could occasionally fail to detect specific attack patterns.

\subsection{Feature Groups and Descriptions}

\textcolor{blue}{To better understand the behavior of different input variables in the context of DoS attack detection, we grouped the extracted features into several logical categories based on their functional characteristics. This breakdown provides insight into which types of features are more indicative of attack patterns versus those that tend to resemble benign traffic. Below, we describe each feature group along with an analysis of its potential relevance and limitations in identifying DoS traffic.}

\subsubsection{Flow duration and packet lengths}

\begin{table}[h]
\color{blue}
\small
\centering
\caption{Flow duration and packet length features.}
\label{tab:flow-duration-pck-len}
\begin{tabular}{ll}
\toprule
\textbf{Feature Name} & \textbf{Description} \\
\midrule
$flow\_duration$ & Total duration of the network flow \\
$pkt\_len\_max$ & Maximum packet length in the flow \\
$pkt\_len\_min$ & Minimum packet length in the flow \\
$pkt\_len\_mean$ & Mean (average) packet length \\
$pkt\_len\_var$ & Variance in packet lengths \\
\bottomrule
\end{tabular}
\label{tab:flow-duration-features}
\end{table}

\textcolor{blue}{Table \ref{tab:flow-duration-pck-len} highlights the flow duration and packet length features, which} measure the overall duration of network flows and various packet length statistics. While these metrics are generally informative, they may not exhibit significant changes during a DoS attack, which typically involves a high volume of short, repetitive requests rather than substantial variations in flow durations or packet lengths.

\subsubsection{Throughput metrics}

\begin{table}[h]
\color{blue}
\footnotesize
\centering
\caption{Throughput metrics features.}
\label{tab:through-metrics}
\begin{tabular}{ll}
\toprule
\textbf{Feature Name} & \textbf{Description} \\
\midrule
$bytes\_per\_s$ & Number of bytes transmitted per second \\
$pkt\_per\_s$ & Number of packets transmitted per second \\
$fwd\_pkt\_per\_s$ & Number of forward packets transmitted per second \\
\bottomrule
\end{tabular}
\label{tab:throughput-features}
\end{table}

\textcolor{blue}{Table \ref{tab:through-metrics} lists the throughput metrics features, which} represent the network throughput in terms of bytes and packets per second. In the context of DoS attacks, particularly volumetric attacks, these throughput metrics may not serve as effective differentiators, as attack traffic can closely mimic normal high-volume traffic patterns.

\subsubsection{Forward and backward packet lengths}

\begin{table}[h]
\color{blue}
\centering
\footnotesize
\caption{Forward and backward packet length features.}
\label{tab:fwd-bwd-pck-len}
\begin{tabular}{ll}
\toprule
\textbf{Feature Name} & \textbf{Description} \\
\midrule
\multicolumn{2}{l}{\textbf{Forward Packet Features}} \\
$fwd\_pkt\_len\_tot$ & Total length of forward packets \\
$fwd\_pkt\_len\_max$ & Maximum length of forward packets \\
$fwd\_pkt\_len\_min$ & Minimum length of forward packets \\
$fwd\_pkt\_len\_std$ & Standard deviation of forward packet lengths \\
$fwd\_pkt\_hdr\_len\_tot$ & Total header length of forward packets \\
$fwd\_non\_empty\_pkt\_cnt$ & Count of non-empty forward packets \\
\midrule
\multicolumn{2}{l}{\textbf{Backward Packet Features}} \\
$bwd\_pkt\_cnt$ & Total number of backward packets \\
$bwd\_pkt\_len\_tot$ & Total length of backward packets \\
$bwd\_pkt\_len\_max$ & Maximum length of backward packets \\
$bwd\_pkt\_len\_min$ & Minimum length of backward packets \\
$bwd\_pkt\_hdr\_len\_tot$ & Total header length of backward packets \\
$bwd\_pkt\_hdr\_len\_min$ & Minimum header length of backward packets \\
$bwd\_non\_empty\_pkt\_cnt$ & Count of non-empty backward packets \\
\bottomrule
\end{tabular}
\label{tab:packet-length-features}
\end{table}

\textcolor{blue}{Table \ref{tab:fwd-bwd-pck-len} lists the forward and backward packet length features which} represent detailed statistics of packet lengths and counts in both forward and backward directions. Similar to previous packet length metrics, these features may not exhibit distinctive patterns that differentiate DoS traffic from legitimate traffic, particularly when the attack traffic is designed to closely resemble normal traffic.

\subsubsection{Inter-arrival times (IAT) and activity periods}

\begin{table}[h]
\color{blue}
\centering
\footnotesize
\caption{IAT and activity/idle period features.}
\label{tab:IAT-activityIP}
\begin{tabular}{ll}
\toprule
\textbf{Feature Name} & \textbf{Description} \\
\midrule
\multicolumn{2}{l}{\textbf{IAT Features}} \\
$iat\_min$ & Minimum inter-arrival time between any two packets \\
$fwd\_iat\_tot$ & Total inter-arrival time for forward packets \\
$fwd\_iat\_std$ & Standard deviation of forward packet IAT \\
$bwd\_iat\_max$ & Maximum inter-arrival time for backward packets \\
$bwd\_iat\_mean$ & Mean inter-arrival time for backward packets \\
$bwd\_iat\_std$ & Standard deviation of backward packet IAT \\
\midrule
\multicolumn{2}{l}{\textbf{Activity and idle period features}} \\
$active\_max$ & Maximum duration of active periods \\
$active\_mean$ & Mean duration of active periods \\
$active\_std$ & Standard deviation of active durations \\
$idle\_max$ & Maximum duration of idle periods \\
$idle\_min$ & Minimum duration of idle periods \\
$idle\_mean$ & Mean duration of idle periods \\
$idle\_std$ & Standard deviation of idle durations \\
\bottomrule
\end{tabular}
\label{tab:iat-activity-idle-features}
\end{table}

\textcolor{blue}{Table \ref{tab:IAT-activityIP} reveals the IAT and activity/idle period features, which} represent various aspects of timing between packets, as well as periods of activity and idleness. While these inter-arrival time (IAT) and activity metrics can be informative in certain contexts, they may not effectively capture the rapid and overwhelming characteristics of DoS attacks, particularly as attack traffic may not exhibit clear timing patterns that distinguish it from normal traffic, especially in sophisticated scenarios.

\subsubsection{TCP/IP flags}

\begin{table}[h]
\color{blue}
\footnotesize
\centering
\caption{TCP/IP flag features (i.e., flag indicators).}
\label{tab:tcp-ip-flag}
\begin{tabular}{ll}
\toprule
\textbf{Feature Name} & \textbf{Description} \\
\midrule
$flag\_ack$ & Presence of the ACK (Acknowledgment) flag \\
$flag\_psh$ & Presence of the PSH (Push) flag \\
$bwd\_flag\_psh$ & Presence of the PSH flag in backward packets \\
$flag\_cwr$ & CWR (Congestion Window Reduced) flag \\
$flag\_ece$ & ECE (Explicit Congestion Notification Echo) flag \\
\bottomrule
\end{tabular}
\label{tab:tcp-flag-features}
\end{table}

\textcolor{blue}{Table \ref{tab:tcp-ip-flag} highlights the TCP/IP flag features, which} represent the counts of specific TCP/IP flags in network traffic. While analyzing these flags can occasionally help detect DoS attacks, these particular flags may not be as indicative as others, such as SYN or RST flags, in identifying malicious activity.

\subsubsection{Bulk transfer metrics}

\begin{table}[h]
\color{blue}
\footnotesize
\centering
\caption{Bulk transfer features.}
\label{tab:bulk-transfer}
\begin{tabular}{ll}
\toprule
\textbf{Feature Name} & \textbf{Description} \\
\midrule
\multicolumn{2}{l}{\textbf{Forward bulk transfer metrics}} \\
$fwd\_bulk\_bytes\_mean$ & Mean nbr. of bytes in forward bulk transfers \\
$fwd\_bulk\_pkt\_mean$ & Mean nbr. of packets in forward bulk transfers \\
$fwd\_bulk\_rate\_mean$ & Mean rate of forward bulk data transfer \\
\midrule
\multicolumn{2}{l}{\textbf{Backward bulk transfer metrics}} \\
$bwd\_bulk\_bytes\_mean$ & Mean nbr. of bytes in backward bulk transfers \\
$bwd\_bulk\_pkt\_mean$ & Mean nbr. of packets in backward bulk transfers \\
$bwd\_bulk\_rate\_mean$ & Mean rate of backward bulk data transfer \\
\bottomrule
\end{tabular}
\label{tab:bulk-transfer-features}
\end{table}

\textcolor{blue}{Table \ref{tab:bulk-transfer} lists the bulk transfer features, which} represent average metrics related to bulk data transfers in both forward and backward directions. These metrics may not be significantly impacted by DoS attacks, which typically focus on overwhelming the target with numerous small requests rather than large bulk data transfers.

\subsubsection{Subflow metrics}

\begin{table}[h]
\footnotesize
\color{blue}
\centering
\caption{Subflow features.}
\label{tab:subflow}
\begin{tabular}{ll}
\toprule
\textbf{Feature Name} & \textbf{Description} \\
\midrule
\multicolumn{2}{l}{\textbf{Forward subflow metric}} \\
$fwd\_subflow\_bytes\_mean$ & Mean nbr. of bytes per forward subflow \\
\midrule
\multicolumn{2}{l}{\textbf{Backward subflow metrics}} \\
$bwd\_subflow\_bytes\_mean$ & Mean nbr. of bytes per backward subflow \\
$bwd\_subflow\_pkt\_mean$ & Mean nbr. of packets per backward subflow \\
\bottomrule
\end{tabular}
\label{tab:subflow-features}
\end{table}

\textcolor{blue}{Table \ref{tab:subflow} summarizes the subflow features, which} represent the average metrics of subflows in both forward and backward directions. Similar to bulk transfer metrics, these subflow metrics may not exhibit significant changes during DoS attacks.

The common thread among these features is that they measure aspects of network traffic that may not exhibit significant deviations during DoS attacks, especially if the attack traffic is designed to closely mimic legitimate traffic. DoS attacks typically focus on overwhelming the target with high volumes of traffic or specific types of requests (such as SYN floods), rather than altering packet sizes, flow durations, or bulk transfer patterns. Therefore, features that directly capture the volume, frequency, and abnormal flags (like SYN or RST) tend to be more indicative of DoS attacks

The importance of specific features in detecting DoS attacks on a network lies in their ability to capture the abnormal traffic patterns and behaviors characteristic of such attacks. The most significant feature, `bwd\_pkt\_len\_mean', which refers to the mean length of packets traveling from the server to the client, is crucial as DoS attacks often generate large volumes of traffic that can alter the typical packet size distribution. Similarly, `fwd\_pkt\_hdr\_len\_min' and `fwd\_iat\_min', which measure the minimum header length of forward packets and the minimum inter-arrival time between forward packets, respectively, help identify unusual spikes in traffic and rapid packet exchanges that are typical in flood-based attacks. The presence of flags such as `flag\_fin', `flag\_psh', and `fwd\_flag\_psh' is vital as these TCP flags can indicate connection termination or push functions, which can be exploited during DoS attacks to disrupt normal communication. 

Furthermore, `pkt\_len\_var', indicating packet length variance, helps in detecting inconsistencies in packet sizes, another hallmark of a network under attack. Metrics like `fwd\_tcp\_init\_win\_bytes' and `bwd\_tcp\_init\_win\_bytes', representing the initial TCP window sizes, can reveal congestion or unusual usage patterns, as attackers may attempt to manipulate these windows to overwhelm the network. Features such as `active\_min' and `bwd\_iat\_mean' provide insights into the duration and timing of connections, with irregular values pointing to potential attacks. The inclusion of throughput-related features like `bwd\_pkt\_per\_s' and `pkt\_s' is also significant, as DoS attacks typically involve high packet rates intended to saturate the network bandwidth. These features provide a comprehensive view of the network's operational metrics, enabling the detection of deviations from normal behavior that signal the presence of DoS attacks.

\section{Learned Lessons and Future Directions} \label{sec:discussion}

Throughout the development and evaluation of this work, several critical lessons were learned regarding the effective use of automated ML tools and explainable AI in the detection of DoS/DDoS attacks. One of the key insights gained was the value of leveraging automated ML frameworks, such as TPOT, to streamline the model selection and feature optimization processes. Automating these processes reduced the need for manual experimentation, resulting in significant time savings and ensuring that optimal hyperparameters and feature sets were systematically identified. However, it was also evident that the model's success heavily depended on careful feature engineering, as not all features contributed equally to the final prediction outcomes.

The automated model selection process using TPOT identified a DecisionTreeClassifier as the most effective model for DDoS detection, optimizing for both accuracy and computational efficiency. The SHAP analysis revealed that bwd\_pkt\_len\_mean, \\fwd\_pkt\_hdr\_len\_min, and flag\_fin are the most important features contributing to the model's predictions. The selection of these features is justified by their ability to capture critical aspects of network traffic that are indicative of DDoS attacks. For example, abnormal patterns in packet lengths and TCP flags are well-known indicators of various types of DDoS attacks, such as volumetric attacks and protocol-based attacks . The high SHAP values associated with these features confirm their importance in the detection process.

Another lesson learned involved the integration of SHAP to improve the interpretability of ML models. By providing insights into the contributions of individual features to the model's predictions, SHAP not only addressed the ``black box" nature of complex models but also allowed for greater trust and transparency in the decision-making process. This transparency is crucial for cybersecurity applications, where the ability to explain and justify model predictions can be essential for operational decisions. Nonetheless, the interpretability provided by SHAP can be computationally intensive, especially for larger datasets, which may limit its applicability in real-time scenarios without further optimization.

Further, the interpretability provided by SHAP allows network administrators to understand the decision-making process of the model, making it easier to trust and deploy in real-world environments. This is particularly important in operational security settings, where the ability to explain why a certain traffic flow was flagged as malicious is crucial for incident response and mitigation

From the experimental results, it also became apparent that certain features, such as packet length statistics and TCP flags, are particularly influential in detecting DDoS attacks, whereas metrics related to bulk data transfers and subflow measurements were less significant. This finding suggests that future efforts should focus on enhancing feature selection mechanisms to further refine the detection models and potentially improve their efficiency by excluding less relevant features.

Moving forward, several research directions could further enhance the effectiveness and applicability of the proposed methodology. One potential avenue is the integration of real-time SHAP optimization methods to reduce computational overhead and enable the deployment of explainable AI in time-sensitive environments. Developing lightweight interpretability mechanisms that can provide near real-time explanations without compromising accuracy would be invaluable for practical implementations in cybersecurity.

Another promising research direction is to explore the use of deep learning models in conjunction with TPOT to assess whether more sophisticated models, such as Convolutional Neural Networks (CNNs) or Long Short-Term Memory (LSTM) networks, could provide improved detection performance. Such models could be trained to detect more complex patterns in network traffic, potentially improving the accuracy of DDoS detection systems even in the face of evolving attack strategies.

Moreover, expanding the scope of this research to include other types of cyber threats, such as ransomware or Advanced Persistent Threats (APTs), would provide a more comprehensive cybersecurity solution. The adaptability of TPOT to different attack types and the potential for incorporating ensemble learning techniques could yield a more generalized model capable of addressing a wide range of threats in diverse environments.

\section{Conclusion} \label{sec:conclusion}

In this study, we presented a novel approach for detecting Distributed Denial of Service (DDoS) attacks by leveraging the Tree-based Pipeline Optimization Tool (TPOT) for automated machine learning (ML) and SHapley Additive exPlanations (SHAP) for model interpretability. The integration of TPOT significantly streamlined the model selection and feature optimization processes, while SHAP provided essential insights into the contributions of individual features, thereby addressing the challenge of model transparency. Our experimental results demonstrated that features such as mean backward packet length and minimum forward packet header length were critical for effective detection, while others, such as bulk transfer metrics, contributed less to the model's predictive performance.

The proposed framework not only enhances detection accuracy but also ensures that the decision-making process remains interpretable and trustworthy-an essential requirement in cybersecurity applications. Future research will focus on optimizing the scalability and efficiency of the interpretability mechanism, as well as extending the applicability of the framework to detect a broader range of cyber threats, further contributing to robust and transparent network security solutions.

\section*{Author Contributions} 
\textbf{Paul Badu Yakubu:} Conceptualization,  Investigation, Software, Methodology, Data curation, Writing – original draft; \textbf{Lesther Santana:} Investigation, Validation, Formal analysis, Visualization; \textbf{Mohamed Rahouti:} Supervision, Project administration, Writing – review \& editing, Funding acquisition; \textbf{Yufeng Xin: } Validation, Visualization; \textbf{Abdellah Chehri:} Methodology, Validation, Writing – review \& editing; Funding acquisition; Mohammed Aledhari: Validation, Visualization.

\section*{Data Availability Statement} 
The data that support the findings of this study are available from the corresponding author upon reasonable request.

\section*{Conflicts of Interest} 
The authors declare no conflicts of interest.

\section*{funding information} 
Funding Agency: Natural Sciences and Engineering Research Council of Canada (Grant Number: RGPIN-2022-3256).
\bibliographystyle{elsarticle-harv} 
\bibliography{references}

\end{document}